\documentclass{aastex62}

\graphicspath{{./}{figures/}}

\accepted{Nov 6, 2019}
\submitjournal{AJ}

\shorttitle{}
\shortauthors{Astudillo et al.}

\usepackage{amsfonts, amsmath, amssymb}

\usepackage{xcolor}

\begin{document}

\title{An Information Theory Approach on Deciding Spectroscopic Follow Ups}

\correspondingauthor{Javiera Astudillo}
\email{jfastudillo@uc.cl}

\author{Javiera Astudillo}
\affiliation{Computer Science Department, School of Engineering, Pontificia Universidad Cat\'olica de Chile}
\author{Pavlos Protopapas}
\affiliation{Institute for Applied Computational Science, Harvard University, Cambridge, MA, USA}
\author{Karim Pichara}
\affiliation{Computer Science Department, School of Engineering, Pontificia Universidad Cat\'olica de Chile}
\affiliation{Millennium Institute of Astrophysics, Chile}
\affiliation{Institute for Applied Computational Science, Harvard University, Cambridge, MA, USA}
\author{Pablo Huijse}
\affiliation{Informatics Institute, Universidad Austral de Chile, Valdivia, Chile}
\affiliation{Millennium Institute of Astrophysics, Chile}

\begin{abstract}
Classification and characterization of variable phenomena and transient phenomena are critical for astrophysics and cosmology. These objects are commonly studied using photometric time series or spectroscopic data. Given that many ongoing and future surveys are in time-domain and given that adding spectra provide further insights but requires more observational resources, it would be valuable to know which objects should we prioritize to have spectrum in addition to time series. We propose a methodology in a probabilistic setting that determines a-priory which objects are worth taking spectrum to obtain better insights, where we focus “insight” as the type of the object (classification). Objects for which we query its spectrum are reclassified using their full spectrum information. We first train two classifiers, one that uses photometric data and another that uses photometric and spectroscopic data together. Then for each photometric object we estimate the probability of each possible spectrum outcome. We combine these models in various probabilistic frameworks (strategies) which are used to guide the selection of follow up observations. The best strategy depends on the intended use, whether it is getting more confidence or accuracy. For a given number of candidate objects (127, equal to $5\%$ of the dataset) for taking spectra, we improve 37\% class prediction accuracy as opposed to 20$\%$ of a non-naive (non-random) best base-line strategy. Our approach provides a general framework for follow-up strategies and can be extended beyond classification and to include other forms of follow-ups beyond spectroscopy.

\end{abstract}

\keywords{methods: data analysis, methods: statistical -- stars: variables}

\section{Introduction}

Variable phenomena have been astronomical objects of great interest as they reveal us important information about our Universe \citep{Groenewegen2018}. A couple of such examples are: RR Lyrae, which are used to trace distances within our galaxy allowing us to improve our understanding of the Milky Way structure and evolution \citep{Minniti2016}, Mira variables, long-period red giants which correspond to a late-stage phase in the evolution of stars like our Sun \citep{Perrin2004}, Cepheids which are used as distance indicators with well studied physical properties \citep{Tanvir1999, Freedman2001}, Supernovae, explosive events some of which act as standard candles in the cosmological distance scale and have been key in recent discoveries related to Dark Energy \citep{Riess1998, Perlmutter1999, Hicken2009} and Quasars/Active Galactic Nuclei (AGN), static transients which help us understand their host galaxies nature \citep{Nolan2001} and early stages of galaxy formation \citep{Eilers2018}. The classification of transients and the identification of novel variable phenomena is critical for astrophysics and cosmology research. This is reflected in the scientific objectives of past, current and near-future wide-field time-domain surveys such as Supernova Legacy Survey \citep[SNLS,][]{Astier2006, Perrett2010}, ESSENCE \citep{Miknaitis2007},
the Large Synoptic Survey Telescope \citep[LSST,][]{Ivezic2008}, the Sloan Digital Sky Survey-II \citep[SDSS-II,][]{Frieman2008,Sako2008}, the Square Kilometre Array \citep[SKA,][]{Lazio2009}, The Catalina Real-Time Transient \citep[CRTS,][] {Djorgovski2011}, the Dark Energy Survey \citep[DES,][]{Bernstein2012, Abbott2018}, Palomar Transient Factory \citep[PTF,][]{Surace2015} which transitioned to Zwicky Transient Facility \citep[ZTF,][]{Smith2014} and the Panoramic Survey Telescope and Rapid Response System \citep[Pan-STARRS1,][]{Chambers2016, Scolnic2018}. From the aforementioned, ongoing surveys such as ZTF \citep{Smith2014} and future ones like LSST \citep{Ivezic2008} will provide extensive datasets, with an estimated number of alerts per night equal to 0.1 million for ZTF \citep{Masci2019} and 10 million for LSST \citep{LSST2014} from which variable phenomena have to be automatically detected.

Photometry from time-domain surveys allows to detect time variable phenomena such as explosions, accretion, pulsations, eclipses and relativistic phenomena undetectable by other means \citep{Djorgovski2011, Kessler2015}. The emerging of synoptic sky surveys which scan sky areas repeatedly have leveraged time-domain astronomy in the late years \citep{Djorgovski2012}, so that recent and future research for variable phenomena discovery and classification of different real transients have focused on automatic methods that mostly rely on photometry
\citep{Debosscher2007, Richards2011, Bloom2012, Pichara2012, Pichara2013, Mackenzie2016, Castro2018, Martinez-Palomera2018}.
    
Apart from photometry, spectroscopic data provide information such as physical properties; gravity, temperature, chemical compositions and radial velocities which are hardly obtainable otherwise \citep{massey2013}. Spectroscopic surveys usually target objects selected from photometric surveys and often their main purpose is to obtain redshift \citep{Djorgovski2013}. Such examples are:  The Baryon Oscillation Spectroscopic Survey \citep[BOSS,][] {Dawson2013} and The Large Sky Area Multi-object Fiber Spectroscopic Telescope \citep[LAMOST,][] {Luo2015}. In addition to redshift estimation, spectra become helpful  for object identification. Such an example would be the type Ia supernovae (SNe Ia), that play a role as distance indicator and are distinguishable from other kind or supernovae, mainly by the lack of hydrogen lines in their spectra and the presence of pronounced silicon features \citep{campbell2013}. Another example are quasars/active galactic nuclei (AGN) among the brightest objects which help us understand their host galaxies nature \citep{Nolan2001} and early stages of galaxy formation \citep{Eilers2018}. The most certain way to confirm these object types and redshift is through the analysis of their spectroscopy \citep{peters2015}. Lastly, RR Lyraes may be distinguished from Eclipsing Binaries through their spectra jointly with their light curves \citep{Kinman2010}. An spectrum with a resolution of 1 \AA\ can provide enough information to distinguish between a pulsating variable and an eclipsing binary even at low amplitudes.
    
Although photometry usually is the first available resource to inspect an object, the spectra ultimately let us understand their physical properties. It is consequently beneficial to have both to achieve better insights. Unfortunately, spectra use much more resources (observational time) hence, they are infrequently available. For instance, the Sloan Digital Sky Survey \citep{york2000}, an ongoing photometric and spectroscopic survey has imaged nearly 1 billion objects, while for spectra it has around 4 million objects\footnote{For further details regarding SDSS data statistics see \href{https://www.sdss.org/dr15/scope/}{SDSS Scope page}.}. Because of this reason spectroscopic follow-up strategy remains a challenge primarily because of the rise of the data stream from imaging surveys along with potential new interesting objects to study in depth. Assuming that we mostly have photometric data and that we can achieve better prediction of the object's class by adding spectra at a specific cost, which objects should we prioritize? 

In this paper, we propose a model that efficiently finds the celestial objects for which obtaining spectra could improve classification results, with regard to either accuracy or confidence. Each unique object in the dataset is classified, using its time series, into one of the classes with an associated level of confidence. Classification of a given object may be poor either because the classification is wrong (wrong class) or because you have low confidence in the classification. For each object, if its spectrum is queried and used alongside its time series to reclassify the object, either or both the class or the confidence may change. Note that we aim to use the full spectral information additional to time series information to make more informed classifications. Our model helps us to find the objects for which querying for its spectra helps to change to the \textit{true} class or increase its confidence. Similarly, it avoids objects which we either know, with high confidence, what their class is or there is no chance in improving.

Our model assigns a priority to each object by measuring the \textit{information gain} and \textit{indicator} (of the classification change) outcome caused by the addition of a spectrum to already available time series. For this, we first extract features from objects (in both training and test set) for which both time series and spectrum are available. For spectrum features, we first learn automatic feature extraction using autoencoders (Section \ref{VAE}) while for time series we use a set of already existent features. We then train classifiers on time series only and on time series and spectra combined. Then we leave the spectra aside and estimate the spectrum for each object given only its time series. For each object, we estimate the \textit{information gain} and an \textit{indicator} between the estimated spectra jointly with time series with just the time series. We assign a priority to each object using the latter estimators, and a subset of the objects with the highest priority is selected for follow-ups. Objects in the subset are classified according to their time series and real spectrum features whereas non-selected objects are classified using only time series features. Our approach is mainly focused on variable phenomena catalogs although it could be applied to any objects that have: available time series, and potentially available spectra. Note that for at least a subset of them we need both spectra and time series for the training phase.

Section \ref{section:related work} presents related work. Section \ref{section:background theory} describes general background theory. Section \ref{section:problem description} defines how we assess the \textit{information gain} and the \textit{indicator}, Section \ref{section:method description} is dedicated to the proposed method. In Section \ref{section:data} we summarize the data used for the experiments. Section \ref{section:implementation} mentions the libraries and hardware used to implement the different models. Section \ref{section:experimental results} presents the results from the experiments and further work. Finally, Section \ref{section:conclusions} presents overall conclusions and future work.

\section{Related work}
\label{section:related work}

In this section, we describe related works and compare them to ours. First, we present the work of \cite{peters2015}, which combines different types of data. Following we describe the spectra follow-up strategy of \cite{2019Ishida}, which selects objects for labeling to efficiently train a classifier. Next, we show the design of experiment approach mentioned in \cite{Yang2015} work, which deals with the optimal set of filters for estimating the true spectral energy distributions (SED). Lastly, we describe \cite{Xia2016} method that optimizes decision-making process regarding which objects to observe given limited telescope time.

\cite{peters2015} work aims to combine different types of data to enhance classification. They compare using band color features only, time-domain features only, and both for quasar detection task, which implies quasar or non-quasar classification. The color features correspond to \textit{Sloan Sky Digital Survey} \citep[SDSS,][]{york2000} adjacent colors $(u - g,\quad g - r,\quad r - i,\quad i - z)$ while the time series features come from fitting a structure function \citep{Schmidt2010} which characterizes the variability of a time series for each band and object. This is motivated by the fact that variability-based classification misses some quasars at high-redshifts while color-based classification gets highly confused at mid-redshift zones. Their results show that the highest completeness (number of known
quasars correctly classified as quasars divided by the number of known quasars) is reached using both types of features. This is consistent with our motivation for combining data to get a better insight into selected variable phenomena. Similar to our work, they seek to improve classification results by increasing the available information (more features) of the objects and do not improve classification models for the same information. Differently, they use features from broad color bands while we use spectra, which has much more detailed color information.

A relevant work is the spectra follow-up strategy of \cite{2019Ishida} since, as proposed in this work, it selects objects for spectra querying though it was proposed for a different use case. They focus on improving type Ia vs non-Ia SNe photometric classification using Active Learning (AL) \citep{Cohn1996}. AL refers to algorithms that select the best objects to label in order to improve classification prediction the most at each training step. It is thereby a way to improve a classification prediction at minimum labeling cost. In their work, labeling is obtained by means of spectrum confirmation since SNe Ia may be distinguished from non-Ia through some specific spectral features which are linked to a particular physical process \citep{campbell2013}. 

The classifiers obtained through their method can improve \textit{purity} (reduce the falsely predicted SNe Ia) compared to two proposed baseline strategies: \textit{passive learning strategy} and \textit{canonical strategy}. In the \textit{passive learning strategy,} objects to be labeled are randomly selected from the unlabelled set while in the \textit{canonical strategy} they are randomly selected from a sample which closely follows their initially labeled dataset distribution. Similar to our work, they propose a spectroscopic follow-up design strategy which uses active learning strategies to select follow-up objects. The main difference with our work is that we propose to use spectral features to improve the reliability of the prediction while they query for labels (through the spectra) to retrain the classifier to reduce the false positive predictions. We automatically extract features which compress the full spectral information into a set of features so as to increase the available information of each object. This may potentially complement regular classification (by an expert) as the latter entails also some uncertainty. This allows to take full advantage of the spectra and thus more informed decisions when classifying.

The described follow-up strategy could fall in the much broader umbrella of Design of Experiment area \citep{Fisher1935}. Generally speaking, it concerns of which and how one should collect the data for an experiment avoiding wastage of resources or missing of important data. One example of this sort is the work of \citep{Yang2015} which deals with how to choose the optimal set of filters to be used to estimate the true spectral energy distributions (SED) of an astronomical object. Similar to our work, they aim to make efficient use of telescope resources. Differently from us, they optimize the observations (i.e. which and how many filters) that are required to build the SED for a single object while we decide which objects to observe in the first place (i.e. taking the spectra).

The work of \cite{Xia2016} presents a method for selecting objects to observe in a batch fashion optimizing telescope resources use. It takes into account the field of view (FOV) of the telescope in use so that many objects may be observed at the same time. Also, it proposes updating and recommending objects to observe in batches given that the infrastructure in a telescope is not able to update observations schedule in real time but it can do it by batches. They use active learning to select objects and take advantage of the location of the candidates in the sky to reach feasible ways of scoring them. They use other active learning approaches to evaluate the performance of their method. These include: (1) randomly selecting the points, (2) selecting the most uncertain points for the current classifier, (3) selecting the most under-sampled points according to the training set distribution, (4) selecting points that maximized the change in the predicted probabilities, (5) active learning and semi-supervised learning methods from other works. Their approach surpass all of the rest regarding the observing time and number of queries required to reach a given accuracy. Similar to our work they select objects to observe to improve the classifier prediction without wastage of telescope resources. Differently from us, they do not mix different sourced data but instead only use time series features \citep[$FATS$,][]{nun2015} to classify. Furthermore, their focus is not to add information to already observed objects and hence gain better insight but rather improve their classifiers through labeling.

\section{Background theory}
\label{section:background theory}

\subsection{Information Theory and Entropy}
\label{subsection:information theory}

Information theory was conceptualized in \citep{Shannon1948} to solve the problem of optimal information transmission over a noisy channel. In his seminal work, Shannon proposed information entropy as a measure of uncertainty. Initially conceived as the average rate of information produced by a stochastic process, it has been broadly used for quantifying information, choice, and uncertainty. Shannon's entropy is defined as
\begin{equation}
\label{eqn:entropy_original}
    H = - \sum_{i=1}^{n}p_{i} \log\;p_{i},
\end{equation}
where $p_{1}...p_{n}$ are the probabilities associated with a set of possible outcomes correspondingly. For this particular work equation \ref{eqn:entropy_original} will always refer to the entropy of one single object. It can be seen from the definition that the less probable an event is the higher its contribution to entropy, which translates in more information needed to describe those events.

Two of the main properties that make it a suitable measure regarding the amount of information are: (1) $H = 0$ when there is only one certain event and therefore no amount of information is needed to describe it. (2) $H$ is maximum when all $p_i$ are equal (i.e $1/n$) and consequently, the uncertainty on the outcome is maximum so that a higher amount of information is needed to describe it.

\subsection{Autoencoders}
\label{VAE}

Autoencoders \citep{Olshausen1996, Lee2006, Vincent2008} are algorithms that learn data representations automatically. These representations can be used as features for subsequent tasks of classification or clustering. The most basic autoencoder is composed of a deterministic encoding function that maps the input $x$ to a hidden or latent space $z$ and a deterministic decoding function that returns $\hat x$ from $z$. The model is trained by minimizing the error between $\hat x$ and $x$. By imposing that $z$ dimension is much lower than $x$ we force the model to learn the most relevant features of the data. Different techniques exist to enhance the encoding such as sparse autoencoders \citep{Olshausen1996, Lee2006} which through regularization force sparsity on $z$, and denoising autoencoders \citep{Vincent2008} which reconstruct $x$ from a corrupted version of it in order to make the model robust to noise.

Autoencoders have been combined with recurrent neural networks \citep{Hochreiter1997} so as to learn low dimensional representations for time series \citep{srivastava2015, witten2016}. These so-called sequence-to-sequence autoencoders have been developed for dense and regularly sampled time series. Important work related to our work is \cite{Naul2018} where an autoencoder for astronomical time series classification was proposed. A shortcoming of this work is that it performs poorly on non-periodic or unfolded time series.

On the other hand, Variational Autoencoder (VAE) \citep{kingma2013} or alternatively Deep Latent Gaussian Model (DLGM) \citep{Rezende2014} is a deep generative model which resembles an encoder and a decoder structure but instead of looking for a deterministic encoding $z$ or decoding $\hat{x}$ it seeks to estimate their distributions. They provide an amortized model for variational inference (VI) \citep{Blei2016}, using the same model to estimate the variational parameters of different data points, avoiding costly loops per data point. VI is a family of techniques to approximate computationally intractable posterior distributions via solving an optimization problem. In VAE we approximate the posterior of z given x using a factorized Gaussian distribution. The encodings follow a regular geometry (usually a Gaussian distribution) and are more meaningful than the ones obtained through a regular autoencoder. This model may be used for artificial data generation, data representation, and inference tasks. A couple of examples are collaborative filtering \citep{Liang2018} and image analysis \citep{Wang2017}. 

Extensions to VAE for fitting sequential data have been made such as \citep{bowman_2015} for sentence generation, \cite{Fabius2014} for simple video game songs datasets and \cite{Chung2015} for speech and handwriting datasets. Sequence VAE were developed for regularly sampled time series and thereby are not appropriate for astronomical data. It is worth noting that training a sequence VAE is more difficult than conventional VAE. The optimization challenges of sequence VAE are described by \cite{bowman_2015} in their Section 3.1 and \cite{dieng2018} in their Section 4.2.

\section{Problem description and notation}
\label{section:problem description}

Our goal is to select objects for which spectra improve class prediction the most if added to its time series, with regard to either accuracy or confidence. We define two metrics to assess classification improvement. The first one is \textit{information gain} and addresses the confidence. It uses the entropy \citep[Section \ref{subsection:information theory}] {Shannon1948} over some estimator $\hat{y}$ of the label $y$ (the \textit{real} class of a given object):

\begin{equation}
H (\hat{y}| x_{*}) = - \sum_{c \in C}
P(\hat{y} = c | x_{*})
\log P(\hat{y} = c | x_{*}).
\label{eqn: entropy_classification}
\end{equation}

Here $\hat{y}$ is an arbitrary class predictor given certain data $x_*$ and $P$ is the probability of predicting a certain class $c$ (i.e. $\hat{y}=c$), where $c$ is any class from the set of possible classes to predict $C$. For example, $\hat{y}$ could be the outcome of any classification method which outputs class probabilities, so that the prediction for a given object with data $x_{*}$ is the class with the highest probability $P$. Entropy $H$ \citep[][eq. \ref{eqn: entropy_classification}]{Shannon1948} let us measure the confusion of a probability density function (PDF) such as the outcome of a classification task. To calculate entropy we need to check the value of the PDF at each value in the domain, which for our case is each of the possible classes to predict. The higher the value of the entropy, the more confused the outcome is while the lower the value, the less confused the outcome is. A confused outcome is whenever probabilities are more even between them and hence we are uncertain of the class. A non-confused outcome is whenever the probabilities are concentrated in one or few classes so that we are more certain of the class. Note that our definition of $H$ works for any feature $x_{*}$ not only for photometric or spectroscopically features ($x_t$ and $x_s$ correspondingly). This means that our method could include adding any different sourced information as long as features may be extracted from it.

We now define the \textit{information gain} for a given object $x$ as the reduction of entropy in the classification outcome caused by the addition of spectrum features $x_s$ to the initially available time series features $x_t$:

\begin{equation}
    IG(x_t, x_s)
    = H_{t}(\hat{y} | x_t)
    - H_{ts}(\hat{y} | x_t, x_s),
    \label{eqn:entropy_diff}
\end{equation}
where
\begin{equation}
\label{eqn:entropy_def}
\begin{gathered}
    H_{t} (\hat{y}| x_t) = -\sum_{c \in C} 
    P_{t}(\hat{y} = c | x_t)
    \log P_{t}(\hat{y} = c | x_t)
    \\
    H_{ts} (\hat{y}| x_t, x_s) = -
    \sum_{c \in C} 
    P_{ts}(\hat{y} = c | x_t, x_s)
    \log P_{ts}(\hat{y} = c | x_t, x_s).
\end{gathered}
\end{equation}

We have used $H_{ts}$, $H_{t}$ to denote the difference between the distributions of their corresponding class predictors $P_{ts}$, $P_{t}$ depending on the given data. Note that we subtract the resulting entropy from using spectrum features to the initial entropy since we want to measure the entropy reduction or equivalently the confusion reduction.

Additionally, to the \textit{information gain} metric, we develop a second \textit{indicator} metric which addresses the classification accuracy. For a given object, it indicates if the predicted class with time series information is different from the prediction with the spectrum information added. This eases the detection of objects which are wrongly classified (\textit{false positives}) with only time series features $x_t$ but are correctly classified if spectrum features $x_s$ are added. The \textit{indicator} function is as follows:

\begin{align}
\label{eqn:class_change}
\Delta \hat{y}(x_t, x_s) = \left\{ \begin{array}{cc} 
    1, &  l_t(x_t) \neq l_{ts}(x_t, x_s) \\
    0, &  l_t(x_t) = l_{ts}(x_t, x_s) \\
    \end{array} \right.
\end{align}

\begin{equation}
\label{eqn:l_t}
    l_{t}(x_t) = arg\max_{c} P_t(\hat{y}=c|x_t)
\end{equation}

\begin{equation}
\label{eqn:l_ts}
    l_{ts}(x_t, x_s) = arg\max_{c} P_{ts}(\hat{y}=c|x_t, x_s)
\end{equation}
where $l_{t}$ is the class/label predicted given time series data, $x_t$, calculated as the class with the highest probability assigned by the predictor $\hat{y}$. Similarly, $l_{ts}$ is the class predicted given time series and spectrum data, $x_t$ and $x_s$. 

\begin{figure}[h]
\centering
\includegraphics[width=0.4\textwidth]{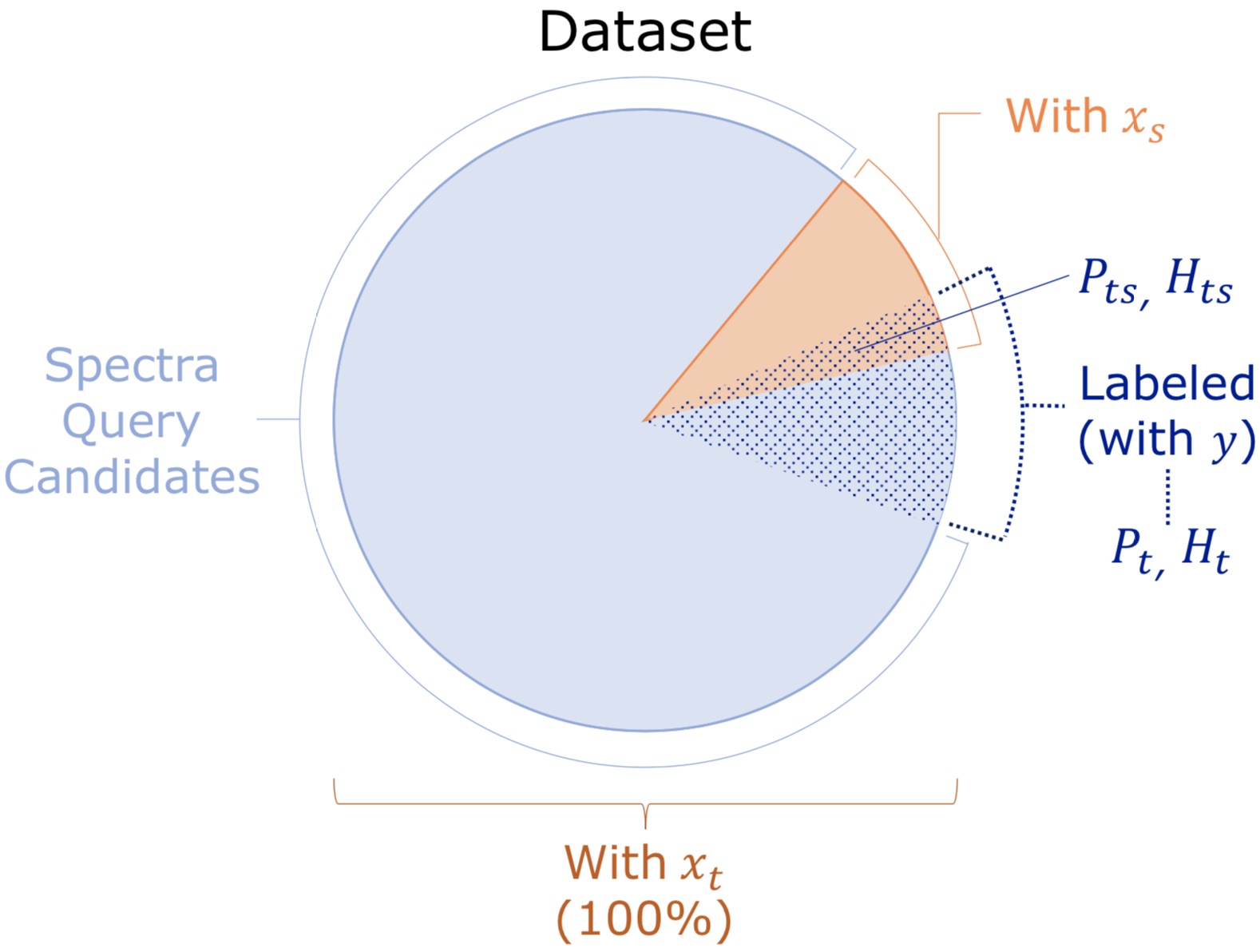}
\caption{Problem description. Time series data ($x_t$) are available for all objects in the dataset while spectroscopic data ($x_s$) for only some of them. A subset of objects have $\{x_t,x_s,y\}$ which are used to estimate $P_{t}$, $P_{ts}$, $H_{t}$ and $H_{ts}$. We want to select objects from ``Spectra Query Candidates'' subset to query for their spectra so as to improve their classification results the most.}
\label{figure:chart}
\end{figure}

In summary, we have time series data, $x_t$, for all objects and spectrum data, $x_s$, for only some of them. We also have a subset of objects which have $\{x_t,x_s,y\}$ which we use to estimate classifiers $P_{t}$, $P_{ts}$ with their corresponding entropy functions, $H_{t}$ and $H_{ts}$ respectively. This is depicted in Figure \ref{figure:chart}. We want to select objects which do not have $x_s$ and have not yet been labeled (unknown $y$) to query for its spectra, so as to improve their classification results the most. These objects are the ones which either improve their classification confidence or are reclassified into their \textit{real} class when their spectra is queried, embodied by functions $IG(x_t, x_s)$ (eq. \ref{eqn:entropy_diff}) and $\Delta \hat{y}(x_t, x_s)$ (eq. \ref{eqn:class_change}) respectively. Both these equations assume we know $x_s$ and require it in their calculations; nevertheless we will evaluate them for objects we have not observed their spectra yet (unobserved $x_s$). Our method focus on how to approximate $x_s$. We propose to replace them with an average over the most probable $x_s$ conditioned over $x_t$. Thus we will need to model the conditional distribution of $x_s$ given $x_t$. 

\section{Method description}
\label{section:method description}

This section describes the methodology for developing the selection criterion. Subsection \ref{subsection:spectra_method} describes how we deal with the fact that $x_s$ are not observed. Subsection \ref{subsection:classifiers} describes the classification models,  $P_{t}$, $P_{ts}$, corresponding to $H_{t}$ and $H_{ts}$, respectively.
Subsection \ref{subsection:features} describes the feature extraction methods for $x_t$, $x_s$. Subsection \ref{subsection:strategies} describes the proposed and comparison strategies. Finally, subsection \ref{overview_method} presents an overview of the proposed method.

\subsection{Unobserved $x_s$} \label{subsection:spectra_method}

Since we do not observe the spectra prior to selecting the candidates, evaluating $H_{ts} (\hat{y}| x_t, x_s)$ from eq. \ref{eqn:entropy_diff} directly is unfeasible. Instead, we propose to replace it with  an average over the most probable $x_s$ conditioned over $x_t$ which in turn will be represented by a learned distribution $Q(x_s|x_t)$ as follows:

\begin{align}
    \label{eqn:h_ts_approx}
    \overline{H}_{ts}( \hat{y}| x_t)
    &= \int H_{ts}(\hat{y}| x_t, x_s) \, Q(x_s| x_t)  dx_s \nonumber \\
    &= \mathbb{E}_{x_s\sim Q(x_s|x_t)}[ H_{ts}(\hat{y}| x_t, x_s)] \nonumber \\
    &\approx \frac{1}{N} \sum_{i=1}^{N} H_{ts} (\hat{y} = y | x_t, \hat{x}_{s,i}),
\end{align}
\begin{equation}
    \hat{x}_{s,i} \sim Q(x_s|x_t).
\end{equation}

We use a Gaussian Mixture Model (GMM) to estimate the joint distribution  $Q(x_t, x_s)$ from which we derive the conditional distribution $Q(x_t| x_s)$ using Bayes rule. We use GMM due to its flexibility to approximate any distribution. We approximate $\overline{H}_{ts}( \hat{y}| x_t)$ through a Markov Chain Monte Carlo (MCMC) process, were $\hat{x}_{s,i}$ is the i-the spectrum sampled from the GMM conditioned on $x_t$.

Replacing $H_{ts}(\hat{y}|x_t, x_s)$ for $\overline{H}_{ts}(\hat{y}|x_t)$ in eq. \eqref{eqn:entropy_diff} the \textit{information gain} function of an object is now:
\begin{equation}
    \label{eqn:info_diff_update}
    IG(x_t) \approx H_{t}(\hat{y}| x_t) -\overline{H}_{ts}(\hat{y}| x_t).
\end{equation}

Similarly for $l_{ts}(x_t, x_s)$ defined in eq. \ref{eqn:l_ts}, the predicted class of $P_{ts}(x_t,x_s)$, we replace it with:
\begin{align}
    \label{eqn:l_ts_approx}
    \overline{l}_{ts}(x_t) 
    &=mode_{x_{s} \sim Q(x_s|x_t)} [l_{ts}(x_t, x_s)] \nonumber\\
    &\approx mode_{\hat{x}_{s,i}} [l_{ts}(x_t, \hat{x}_{s,i})],
\end{align}
\begin{equation}
    \hat{x}_{s,i} \sim Q(x_s|x_t).
\end{equation}

so that eq. \ref{eqn:class_change}, the \textit{indicator} function, is replaced with:
\begin{align}
    \label{eqn:class_change_update}
    \Delta \hat{y}(x_t) = \left\{ \begin{array}{cc} 
    1, &  l_t(x_t) \neq \overline{l}_{ts}(x_t) \\
    0, &  l_t(x_t) = \overline{l}_{ts}(x_t). \\
    \end{array} \right.
\end{align}

The calculation of our updated information gain $IG(x_t)$ and indicator $\Delta \hat{y}(x_t)$ needs the classifiers $P_{t}$ and $P_{ts}$ trained with $x_t \in \mathbb{R}^{t}$ and with $\{x_t, x_s\} \in \mathbb{R}^{t+s}$ correspondingly.

\subsection{$P_{t}, P_{ts}$}
\label{subsection:classifiers}

As mentioned in Section \ref{section:problem description}, $P_{t}$ and $P_{ts}$ may be any type of classifier that output probabilities rather than just hard predictions. The models in this work are based on Random Forest classifiers \citep{breiman1984}. RF classifiers are shown to be a good compromise between performance, efficiency and easy training when features are already extracted.   
$P_{ts}$ is trained over the joint space $\{x_t, x_s\} \in \mathbb{R}^{t+s}$ while $P_t$ is over $x_t \in \mathbb{R}^{t}$ space. The proportion of trees voting for each class is taken as the output probability.

\subsection{$x_t, x_s$}
\label{subsection:features}
In general, the performance of the GMM is inversely proportional to the dimensionality of the features. This is explained by the \textit{curse of dimensionality} \citep{Bellman1961}, a term that refers to the issues that arise when working in high dimensional spaces. As explained in \cite{Bishop2006} a Gaussian distribution probability mass spreads on the tails as the dimensionality increases so that most mass gets concentrated in a thin shell, thereby losing its characteristic shape and becomes unsuitable for some tasks. For this reason, it is more suitable that $x_t$ and $x_s$ are of low dimension. 

$x_t$ is a subset of expert features \citep[FATS,][]{nun2015} while $x_s$ is built from learned features ($\mu(z)$), where \textit{expert} refers to well known studied features and \textit{learned} are those which are automatically extracted, for instance with an autoencoder (subsection \ref{VAE}). $\mu(z)$ is the mean of the latent space $z$ of a VAE trained to encode and decode spectra. Note that we will use a VAE only for spectrum feature extraction and that time series will use a set of already existent features. Spectra need to be preprocessed to have a common input shape for the VAE as described in later subsection \ref{spectra dataset} and depicted in Appendix \ref{appendix: VAE}, Figure \ref{figure:vae_network}. As mentioned in subsection \ref{VAE}, the latent space of a VAE follows a regular geometry which simplifies the GMM modeling. Since we do not know for certain the most suitable dimensionality of the latent space $z$ for our framework we will try different dimensionalities $d$ and select the one which reconstructs the best the spectra over a test set. The test set contains multiple spectra which were not used for training to provide an unbiased evaluation of the model being tested. We train multiple VAE for $d \in [1,15]$, $d \in \mathbb{N}$, $z \in \mathbb{R}^{d}$. The selected $d$ is the one with the least test $R^{2}$ (coefficient of determination) between the original spectra and the reconstructed spectra which is the output of the VAE. As mentioned in sec. \ref{VAE} current extensions of VAE for sequences are both difficult to train and not capable of dealing with unevenly sampled and non-periodic time series. For time series features we opt to use a subset of $FATS$ \citep{nun2015}. 

We select $x_t \subseteq FATS$ and $x_s \subseteq \mu(z)$ jointly to avoid repeated information. We build candidate subsets $C_{d'} \subseteq \{FATS, \mu(z)\}$, $C_{d'} \in \mathbb{R}^{d'}$ with $1 \leq d' \leq |FATS|+|\mu(z)|$. The latter notation refers to the cardinality of the set so that $C_{d'}$ may be a subset of size 1 at the minimum and a subset with a size equal to the number of $FATS$ together with $\mu(z)$ at the maximum. Subset $C_{d'}$ corresponds to the $d'$ most important features (Gini importance) according to a Random Forest classification task as described in \cite{breiman1984}. Note that $x_t$ is not necessarily of the same dimensionality as $x_s$. The selected features are the smallest $C_{d'}$ which improves over 90\% of the possible improvement of the accuracy, between the subset with the worst accuracy and the subset with the best accuracy. The selection is depicted in Figure \ref{fig:overview}.

\subsection{Strategies}
\label{subsection:strategies}
To assign a priority to query for spectrum for each object in the dataset, we develop multiple strategies and compare them with two base-line strategies and one \textit{ideal scenario} strategy. Each strategy assigns a priority to each object and selects a \textit{subset} of objects which to add spectrum features $x_s$ to their available information. For any arbitrary \textit{subset size} ($s$), the $s$ objects with the highest priority are selected for each strategy. Selected objects are classified according to their $x_t$ and $x_s$ features. Non-selected objects are classified using only $x_t$ features. We consider two base-line strategies to compare with ours:

\begin{enumerate}
    \item \textbf{Random}: randomly selects the objects from the dataset.
    
    \item $\mathbf{H_t}$: selects the objects with highest entropy on the time series classification outcome $H_t(\hat{y}|x_t)$. This strategy is similar to common active learning strategies which query for the label such as \cite{2019Ishida}.
\end{enumerate}

We develop three strategies which assign priority to each object according to the estimated information gain $IG(x_t)$ (eq. \ref{eqn:info_diff_update}) and indicator $\Delta \hat{y}(x_t)$ (eq. \ref{eqn:class_change_update}):

\begin{enumerate}
    \setcounter{enumi}{2}
    \item $\mathbf{IG(x_t)}$: selects the objects with highest approximate information gain $IG(x_t)$ (eq. \ref{eqn:info_diff_update}).
    
    \item $\mathbf{IG(x_t) + \Delta \hat{y}(x_t)}$: selects first the objects with $\Delta \hat{y}(x_t) = 1$ (eq. \ref{eqn:class_change_update}) and then according to the highest $IG(x_t)$. This strategy prioritizes the objects whose predicted class using the sampled spectra (eq. \ref{eqn:class_change_update}) is different from that which only uses the time series (eq. \ref{eqn:l_t}). The focus is to detect the objects which are falsely predicted with time series features $x_t$ only ($l_t(x_t) \neq y$) but are correctly predicted if spectrum features $x_s$ are added ($l_{ts}(x_t, x_s) = y$).

    \item $\mathbf{H_t + \Delta \hat{y}(x_t)}$: selects first the objects with $\Delta \hat{y}(x_t) = 1$ (eq. \ref{eqn:class_change_update}) and then according to the highest entropy on the time series classification outcome $H_t(\hat{y}|x_t)$. Similar to the previous strategy, it focuses on detecting falsely predicted objects by combining the base line strategy $H_t$ and our proposed \textit{indicator} (eq. \ref{eqn:class_change_update}). It avoids choosing objects with certain classification (low $H_t(\hat{y}|x_t)$) and lift objects which we think will have a change of prediction if spectrum information is added. 
    
\end{enumerate}

Finally, one ideal scenario for $IG(x_t)$ is included:

\begin{enumerate}
    \setcounter{enumi}{5}
    \item \textbf{Ideal scenario}: selects the objects according to the current value for $IG(x_t, x_s)$ (eq. \ref{eqn:entropy_diff}) using $H_{ts}(\hat{y}|x_t, x_s)$ instead of its approximation $IG(x_t)$ (eq. \ref{eqn:info_diff_update}) which uses $\overline{H}_{ts}(\hat{y}| x_t)$. Note that this is not a feasible strategy for selection (since it uses $x_s$), but a reference of how would $IG(x_t)$ perform if the estimation of the spectra for each objects was the real spectrum.
\end{enumerate}

\subsection{Overview of the method} \label{overview_method}

\begin{figure*}
\includegraphics[width=1\linewidth]{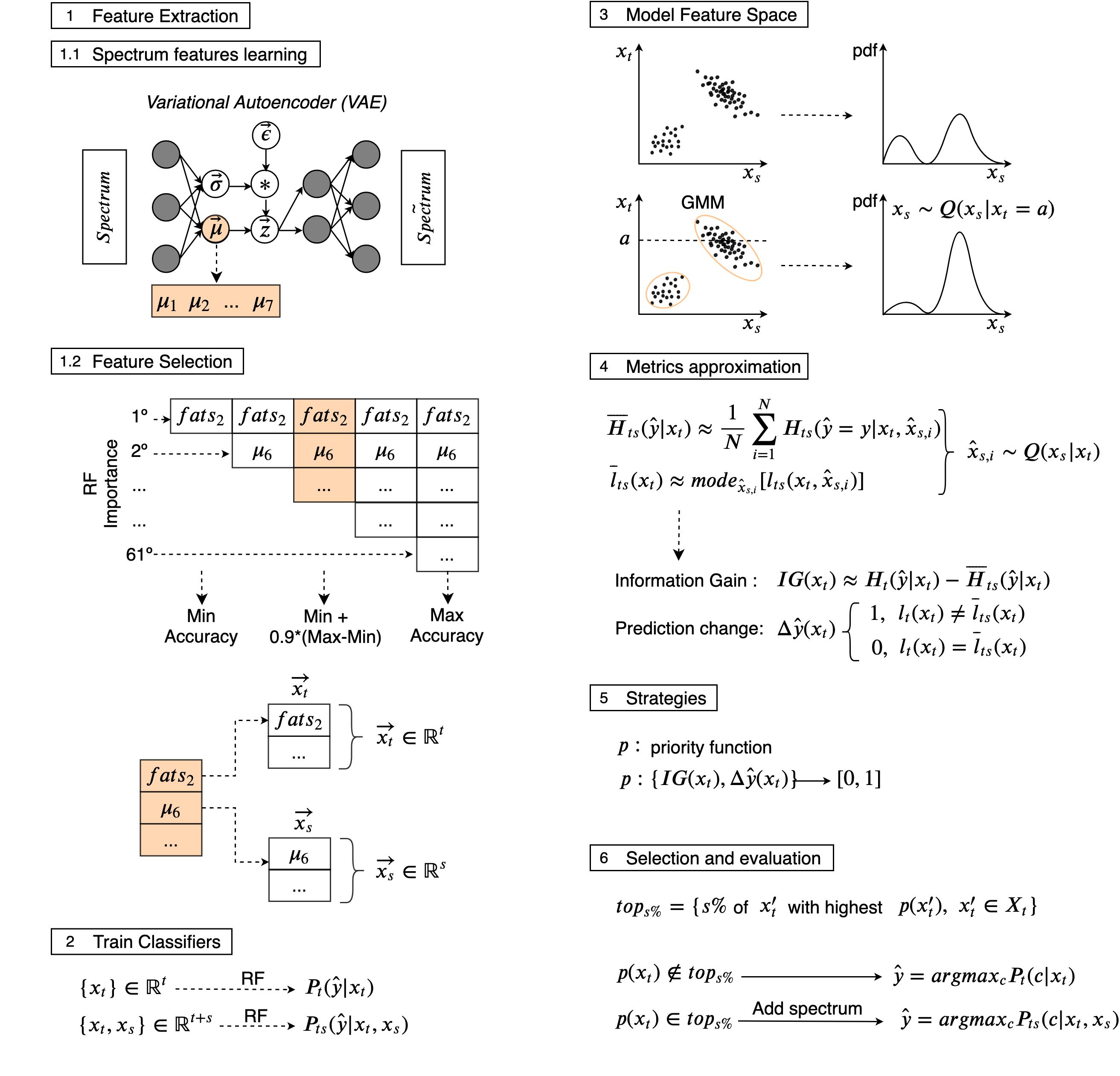}
\caption{Proposed method overview. \textbf{(1)} We first extract time series and spectrum features for all objects in our dataset. \textbf{(2)} We then train classifiers with these features. \textbf{(3)} After that we model the joint space of time series and spectrum features so that \textbf{(4)} spectrum is estimated from its time series features to calculate \textit{information gain} and \textit{indicator} function. \textbf{(5)} Different strategies are proposed to assign a spectrum query priority to each object. \textbf{(6)} Finally, objects with highest priority are selected and classified using their time series and spectrum features while non-selected objects are classified with just time series features.}
\label{fig:overview}
\end{figure*}
 
 Following we give a brief description of our method:
 \begin{enumerate}
     \item \textbf{Feature Extraction}. We extract low dimensional features of time series ($x_t \in \mathbb{R}^{t}$) and spectra ($x_s \in \mathbb{R}^{s}$) for all objects that have both available simultaneously. $x_t$ is a subset of expert features \citep[FATS,][]{nun2015} while $x_s$ is built from learned features ($\mu(z)$). $\mu(z)$ is the mean of the latent space $z$ of a VAE trained to encode and decode spectra. $x_t$ and $x_s$ are jointly chosen as the smallest subset of the most important features according to a Random Forest classification task which achieves over $90\%$ of the maximum improvement between the subsets with the worst and the best accuracy.
     
      \item \textbf{Train Classifiers.} Two Random Forest classifiers are trained over $x_t \in \mathbb{R}^{t}$ space and over the joint space $\{x_t, x_s\} \in \mathbb{R}^{t+s}$. The output classifiers are  $P_{t}(\hat{y}|x_{t})$ and $P_{ts}(\hat{y}|x_t, x_s)$. 
     
     \item \textbf{Model feature space.} An empirical GMM is fitted over the joint space $\{x_t, x_s\} \in \mathbb{R}^{t+s}$ so that $Q(x_{t}, x_{s})$ is the distribution over the joint space. The conditional distribution $Q(x_{s}|x_t)$ is obtained from the GMM properties.
     
     \item \textbf{Metrics approximation.} For each object we evaluate $H_{t}(\hat{y}| x_t)$, $\overline{H}_{ts}(\hat{y}| x_t)$, $l_t(x_t)$, $\overline{l}_{ts}(x_t)$, the entropy and predicted classes of the classifiers trained over the $x_t$ and the $\{x_t, x_s\}$ space, respectively. With these we assess $IG(x_t)$ and $\Delta \hat{y}(x_t)$, the approximate \textit{information gain} and \textit{indicator} function. 
     
     \item \textbf{Strategies.} We assign a priority to each \textbf{object} with calculated metrics through different strategies.
     
     \item \textbf{Selection and Evaluation.} For each strategy, we select objects with highest priority according to a threshold or to a fixed number of candidates. For selected objects we query for their observed spectrum features and classify them according to $P_{ts}(\hat{y}|x_{t}, x_{s})$. Note that in the latter case there is no need to approximate the spectrum features $x_s$ since it is given information. Non-selected objects are classified with $P_{t}(\hat{y}|x_{t})$. 
     
 \end{enumerate}
  The above description is depicted in Figure \ref{fig:overview}.

\section{Data}
\label{section:data}

We perform a crossmatch over two catalogs to find objects with both time series and spectra data. The reached surveys are \textit{Catalina Sky Surveys} \citep[CSS,][]{Larson2003} for time series and \textit{Sloan Sky Digital Survey} \citep[SDSS,][]{york2000} for spectra. Both surveys are better detailed in Sections \ref{time series survey} and \ref{spectra survey}, respectively. The built merged dataset is better detailed in Section \ref{cross-match dataset}.

Additionally to the previous dataset, we gather more spectra to train the VAE for spectrum feature extraction from SDSS. This is due to the insufficient spectra in the cross-match for training this kind of network. The spectra only dataset is better detailed in Section \ref{spectra dataset}.

\begin{table}[ht]
\caption{Datasets Overview}
\centering 
\begin{tabular}{c c c c} 
\hline\hline 
 & Spectra & Time Series & Cross-match \\ [0.5ex]
\hline 
Survey & SDSS & CSS & CSS \& SDSS \\ 
Data Release & All & DR1 & DR1 \& All \\
Source type & Spectra & Photometry & Photometry \& Spectra \\
Labeled & No & Yes & Yes \\
N\textsuperscript{o} of time series & - & 2683 & 2683 \\
N\textsuperscript{o} of spectra & 20,602  (20,949) & - & 3296 \\[1ex] 
\hline 
\end{tabular}
\label{table:dataset} 
\end{table}

\begin{table}[ht]
\centering 
\begin{tabular}{c c c c} 
\hline\hline 
 & CSDR1 & Cross-match \\ [0.5ex]
\hline 
EW & 30,743 & 749 (764)\\
EA & 4,683 & 148 (149)\\
beta Lyrae & 279 & - (7)\\
RRab & 16,797 & 343 (346)\\
RRc & 5,469 & 1193 (1219)\\
RRd & 502 & 78 (81)\\
Blazkho & 223 & - (5)\\
RS CVn & 1,522 & 43 (43)\\
ACEP & 64 & - (1)\\
Cep-II & 124 & - (3)\\
HADS & 242 & - (28)\\
LADS & 7 & - (-)\\
LPV & 512 & - (9)\\
ELL & 143 & - (4)\\
Hump & 25 & - (-)\\
PCEB & 85 & - (23)\\
EA\_UP & 155 & - (1)
\\[1ex] 
\hline 
\textbf{Total} & \textbf{61,575} & \textbf{2554 (2683)}
\end{tabular}
\caption{Datasets Label Distribution. Second column shows time series from CSDR1. Third column shows unique objects in the cross-match between CSDR1 with SDSS DR14. The original number of objects are shown in parenthesis while the objects kept to build the dataset used in our experiments are shown without parenthesis. Labels which represented less than $1\%$ are removed alongside objects which do not present spectrum features $x_s$. Labels are retrieved from \cite{Drake2014}.}
\label{table:dataset distribution}
\end{table}

\subsection{CSS Survey} 
\label{time series survey}

The \textit{Catalina Sky Surveys}\footnote{https://catalina.lpl.arizona.edu} \citep[CSS,][]{Larson2003} is a NASA funded project that searches for Near Earth Objects (NEO) and covers between declination $\delta$ = -75 and +65 degrees \citep{Drake2014}. It started in 2004 and has three telescopes: the Catalina Schmidt Survey (CSS) and the Mount Lemmon Survey (MLS) in Tucson, Arizona and the Siding Spring Survey (SSS) in Siding Spring, Australia. They set fields that tile their observed sky. Photometry is obtained using the aperture photometry program SExtractor (Bertin \& Arnouts, 1996). In this work we use data from their first Data Release \citep[CSDR1,][]{Drake2014} which follows 198 million discrete sources monitored between April 2005 and June 2011 with an average of 250 observations per field and an exposure time of 30 seconds. More specifically we use 47,000 objects from the \textit{Catalina Surveys Periodic Variable Star Catalog} \cite{Drake2014} that were found to be periodic variables. An inspection was held by a single person for the labeling of 112,000 periodic candidates. It consisted of the examination and comparison of the phased time series morphology with known types of periodic variables. The class distribution is shown in column ``CSDR1" in Table \ref{table:dataset distribution}.

\subsection{SDSS Survey}
\label{spectra survey}

The Sloan Digital Sky Survey\footnote{https://www.sdss.org/} \citep[SDSS,][]{york2000} is an ongoing project that started its operation in 2000 and consists of three main stages: SDSS-I/II, SDSS-III and SDSS-IV, each of them composed of multiple surveys. It is headquartered at Apache Point Observatory in south east New Mexico and  Las Campanas Observatory in northern Chile. Two main spectographs are used along its surveys \citep{Smee2013}: SDSS spectrograph\footnote{http://classic.sdss.org/dr7/instruments/spectrographs/index.html} and BOSS spectograph\footnote{https://www.sdss.org/instruments/boss\_spectrograph/}. The SDSS spectrograph contains 640 fibers of 3 arcsec of diameter per plate, it covers along $3800-9200$ \r{A} and has a resolution of 1500 at 3800 \r{A} and 2500 at 9000 \r{A} . On the other hand, BOSS spectrograph contains 1000 fibers of of 2 arcsec of diameter per plate, covers $3600-10,400$ \r{A} and has a resolution of $1560-2270$ in the blue channel, $1850-2650$ in the red channel.

The data used in this work is taken from the Data Release 14 (DR14) part of the fourth phase of SDSS (SDSS-IV) \citep{Blanton2017}. All observations used the 2.5 m Sloan Foundation Telescope \citep{Gunn2006}. All of these spectra share the same wavelength grid spacing but differ in the starting or ending point\footnote{http://www.sdss3.org/dr9/spectro/spectro\_basics.php}.

\subsection{Time Series and Spectra dataset}
\label{cross-match dataset}

There are 2,683 unique objects in the original cross-match of SDSS DR14 spectra and CSDR1 photometry shrunk to 2,554 after initial data preprocessing. All objects have one time series but may have two spectra. Their labels are retrieved from \cite{Drake2014}. The class distribution is shown in column ``Cross-Match" in Table \ref{table:dataset distribution}. The original number per class is shown in parenthesis beside the resulting number after discarding elements. Elements may be discarded either because their class represent less than 1\% of the dataset or because they do not have spectrum features $x_s$. A spectrum may not have spectrum features $x_s$ if its wavelength coverage is shorter than the required by the VAE input. The latter is better detailed in the following subsection \ref{spectra dataset}. The resulting cross-match dataset used in our framework has objects from six different classes: EW (contact binary, $29.3\%$), EA (semi-detached binary, $5.8\%$), RRab (fundamental mode RR Lyrae, $13.4\%$), RRc (first-overtone RR Lyrae, $46.7\%$), RRd (double-mode RR Lyrae, $3.1\%$), RS CVn (RS Canum Venaticorum, $1.7\%$).

\subsection{Spectra dataset}
\label{spectra dataset}

For training the VAE for spectrum feature extraction, 20,949 spectra from SDSS DR14 are retrieved as shown in Table \ref{table:dataset}. A VAE implemented with fully-connected neural network has a fixed input size nevertheless not all spectra have the same wavelength grid as explained in Section \ref{spectra survey}. Hence we have to preprocess the spectra so that they all have a common wavelength grid. For this, we establish a starting and ending wavelength. Each is chosen following the value on the 99th percentile of the sorted starting and ending points from all spectra correspondingly. Any spectrum which has a higher starting point or a lower ending point is dismissed. We keep the same wavelength spacing since it is the same for all entries. The resulting starting and ending wavelength are 3,830 \r{A} and 9,174 \r{A} correspondingly with a grid of size 3,794. The processed dataset has 20,602 spectra which is 1.7\% less than the original one.

\section{Implementation}
\label{section:implementation}

The libraries used to implement our method are Tensorflow\footnote{https://www.tensorflow.org/} and scikit-learn \citep{scikit}. To train the VAE we use a GPU GeForce GTX1080Ti, 11GB and the selected model takes 746 s. (12.43 min.) to train. The RF with ten fold cross validation and GMM are trained with a 2.3GHz dual-core 7th-generation Intel Core i5 processor and take 31.97 s. and 2.27 s., respectively. The sampling (with $N=200$) and calculation of $\overline{H}_{ts}( \hat{y}| x_t)$ from eq. \ref{eqn:h_ts_approx} and $\overline{l}_{ts}(x_t)$ from eq. \ref{eqn:class_change_update} take 437 s. (7.17 min.) using the latter hardware. All code is provided \href{https://github.com/jfastudillo/An-Information-Theory-Approach-On-deciding-Spectroscopic-follow-ups.git}{here}. Data can be found \href{https://drive.google.com/drive/folders/1AVentdOhgknlfCAz8aWOOm_fLU3BuosS}{here}.

\section{Results}
\label{section:experimental results}

This section describes the necessary components $x_s$, $x_t$ and $Q(x_t, x_s)$ to test our methodology and presents the results for the different strategies described in subsection \ref{subsection:strategies}. Subsection \ref{subsection:results_features} details the selected $x_t$ and $x_s$, as described in \ref{subsection:features}. Subsection \ref{subsection:joint_distribution} describes the fitted joint distribution $Q(x_t, x_s)$. Finally, subsection \ref{subsection:candidate_selection} presents a comparative study of the performance for multiple metrics for the different strategies.

\subsection{Features}
\label{subsection:results_features}
For the $x_s$ features we train fifteen VAE models each with a different latent space dimensionality $d \in [1, 15]$, $z \in \mathbb{R}^{d}$, $x_s \subseteq \mu(z)$, as described in \ref{subsection:features}. The training set is the Spectra Dataset described in subsection \ref{spectra dataset}, composed of 20,602 spectra as indicated in Table \ref{table:dataset}. The models are trained over 100 epochs with annealing\footnote{When annealing a VAE, a variable weight is added to the term which pushes the encodings to follow a prior distribution (KL divergence) in the cost function at training time, which starts at 0 and progressively increases to 1 through training epochs. It is used so that the autoencoder first learns how to encode and decode correctly and then to better shape the distributions of the encodings.} \citep{Kirkpatrick1983} over the loss function as done in \cite{bowman_2015}. We preprocess spectra so that they have a common wavelength grid between 3,830 and 9,174 \text{\AA} and normalize the flux to the $[0,1]$ range for each spectrum. We select the model with $\mu(z) \in \mathbb{R}^{7}$ which reports $R^2=0.96$ (coefficient of determination) over the test set between the original spectra and the reconstructed spectra as explained in \ref{subsection:features}. The selected model consists of four encoding fully-connected layers with ReLU \citep{Glorot2010} activations with 2,847, 1,900, 953 and seven units respectively. Symmetrically, the decoder has four layers with an output layer size of 3,794 units and a sigmoid activation at the output\footnote{It is suitable to use sigmoid for this case since inputs are normalized to $[0,1]$ range.}. The final model is depicted in Appendix \ref{appendix: VAE}. 

$x_t$ is a subset of expert features \citep[FATS,][]{nun2015} while $x_s$ is built from learned features ($\mu(z)$), $x_t \subseteq FATS$ and $x_s \subseteq \mu(z)$. They are jointly chosen as the smallest subset of the most important features according to a Random Forest classification task which achieves over $90\%$ of the maximum improvement between the subsets with the worst and the best accuracy. Further details are described in subsection \ref{subsection:features}. To refer to each dimension of $\mu(z)$ we will use notation ${\mu_i}, i = 0,...,6$. The dataset used to select $x_t$ and $x_s$ features is the Cross-Match Dataset (Table \ref{table:dataset}) detailed in subsection \ref{cross-match dataset}. The selected features are $x_t$=\{PeriodLS, Freq1\_harmonics\_amplitude\_0, MedianAbsDev, Q31, FluxPercentileRatioMid35, FluxPercentileRatioMid50, Freq1\_harmonics\_amplitude\_1\} and $x_s$=\{$\mu_2$, $\mu_5$\}. The cumulative relative importances (Gini impurity) of the selected features $x_s$, $x_t$ and $\{x_t,x_s\}$ according to a Random Forest are $0.08$, $0.39$ and $0.46$, respectively. The average accuracy of the Random Forest trained over ten stratified data folds using the selected $\{x_t, x_s\}$ features is $0.87$ which equals to $93\%$ of the max improvement over the accuracy between the worst selection of $x_t$ and $x_s$ (with worst accuracy equal to $0.73$) and the best selection of $x_t$ and $x_s$ (with best accuracy equal to $0.88$).

\subsection{Joint distribution}
\label{subsection:joint_distribution}
A joint distribution of the features $x_s$ and $x_t$ over the Cross-Match Dataset (Table \ref{table:dataset}) is estimated using a GMM. As mentioned in subsection \ref{subsection:features}, $x_s$ by construction of the VAE follows a regular geometry (in this case a single mode Gaussian) but FATS need more modes to fully describe the distribution. Because of this the number of clusters is set equal to the number of classes as an initial guess and the number of components adapts according to the data with a Variational Bayesian method as explained in \cite{Bishop2006} and provided in scikit-learn package \citep{scikit}. The resulting GMM has the same number of components as the initial guess which is six for this case.

\subsection{Candidate Selection}
\label{subsection:candidate_selection}

To assign a query priority to the spectrum of each object in the dataset we propose three strategies ($IG(x_t)$;  $IG(x_t) + \Delta \hat{y}(x_t)$;  $H_t + \Delta \hat{y}(x_t)$) and compare them with two base-line strategies (\textit{Random}; $H_t$) and the \textit{Ideal scenario} strategy. Each strategy assigns a priority to each of the objects and selects a \textit{subset} of them for which to add spectrum features $x_s$ to their available information. For any arbitrary \textit{subset size} ($s$), the $s$ first objects with the highest priority are selected for each strategy. Selected objects (in the \textit{subset}) are classified according to their $x_t$ and $x_s$ features ($l_{ts}(x_t, x_s)$). Non-selected objects are classified using only $x_t$ features ($l_{t}(x_t)$).

All strategies are detailed in subsection \ref{subsection:strategies}. Our developed strategies are based on $IG(x_t)$ (eq. \ref{eqn:info_diff_update}) which prioritizes the objects with the highest approximate \textit{information gain} and $\Delta \hat{y}(x_t)$ (eq. \ref{eqn:class_change_update}) which prioritizes the objects that are most likely to change their classification if the spectrum features were added. The $H_t$ base-line strategy prioritizes the objects with highest entropy on the time series classification outcome while \textit{Random} gives random priorities to the objects. Ideal scenario selects the objects according to their actual \textit{information gain} instead of their approximation $IG(x_t)$. Note that the latter is not a feasible strategy for selection, but it is included to study how would the $IG(x_t)$ strategy perform if the real spectra were used instead of the estimated spectra.

\begin{figure*}[h]
\centering
\includegraphics[width=\textwidth]{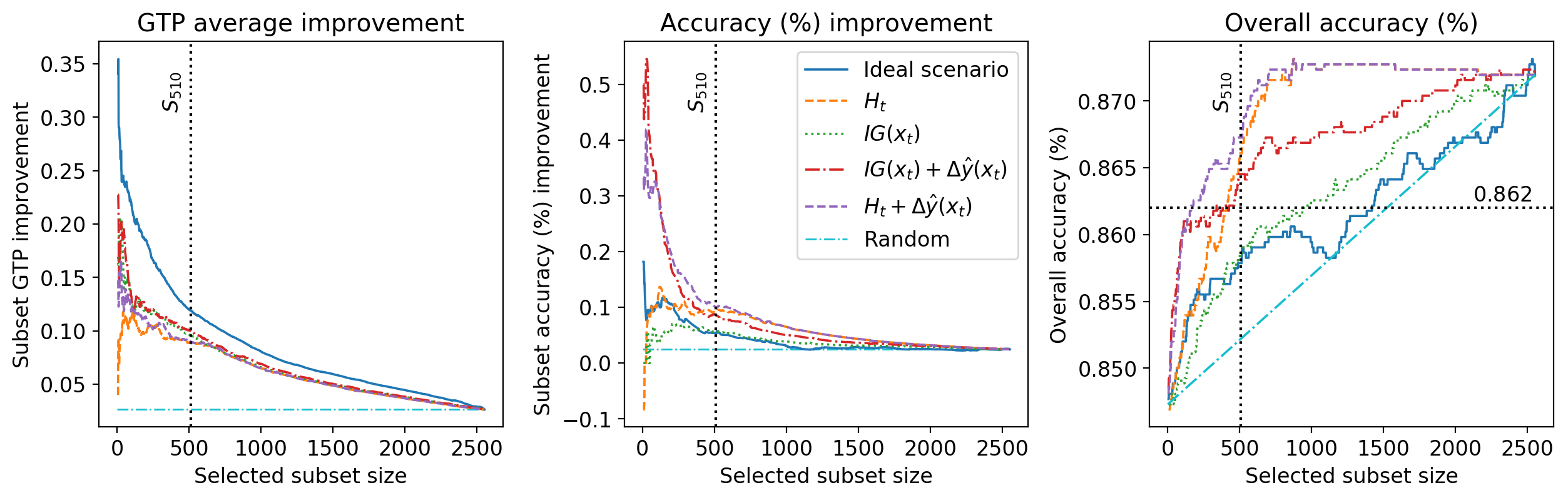}
\caption{Comparative performance of our selection strategies ($IG(x_t)$;  $IG(x_t) + \Delta \hat{y}(x_t)$;  $H_t + \Delta \hat{y}(x_t)$) with base-line strategies (\textit{Random}; $H_t$) and the \textit{Ideal scenario} strategy with respect to \textbf{(Left)} Ground truth probability (GTP) average improvement on selected subset, \textbf{(Middle)} Accuracy (\%) improvement on selected subset and \textbf{(Right)} Overall accuracy. $S_{510}$ represents the subset with 510 objects (equal to $20\%$ of dataset).}
\label{figure:comparison}
\end{figure*}

A comparative performance of the strategies is shown in Figure \ref{figure:comparison}. We evaluate our method using three metrics and different selected \textit{subset size} ($s$) (x-axis). The left plot shows the \textit{subset} average improvement of the Ground Truth Probability (GTP) \textit{i.e.} the average probability assigned to the label $y$ (the \textit{real} class) for each object. Our $IG(x_t) + \Delta \hat{y}(x_t)$ strategy yields the highest GTP average improvement, especially for smaller $s$\footnote{\textit{Ideal scenario} does not compete since it is only a reference strategy and not a feasible one.}. The next better strategy is $IG(x_t)$ which surpass $H_t(x_t) + \Delta \hat{y}(x_t)$ and $H_t$ specially within the first $s$ ($\approx s<S_{510}$) and lastly \textit{Random} strategy is the worst, with a constant average GTP improvement of 0.026. In this scenario, $\Delta \hat{y}(x_t)$ detects 113 objects which are most probable to have a change of classification prediction (with $\Delta \hat{y}(x_t)=1$). Strategies which use \textit{indicator} $\Delta \hat{y}(x_t)$ lift up all objects with $\Delta \hat{y}(x_t)=1$ to have the highest priority so that they are the first to be selected for spectra querying. This way the performance of the latter strategies in any regard is affected by this \textit{indicator} only up to $s=113$. 

\begin{figure}[h]
\centering
\includegraphics[width=1\textwidth]{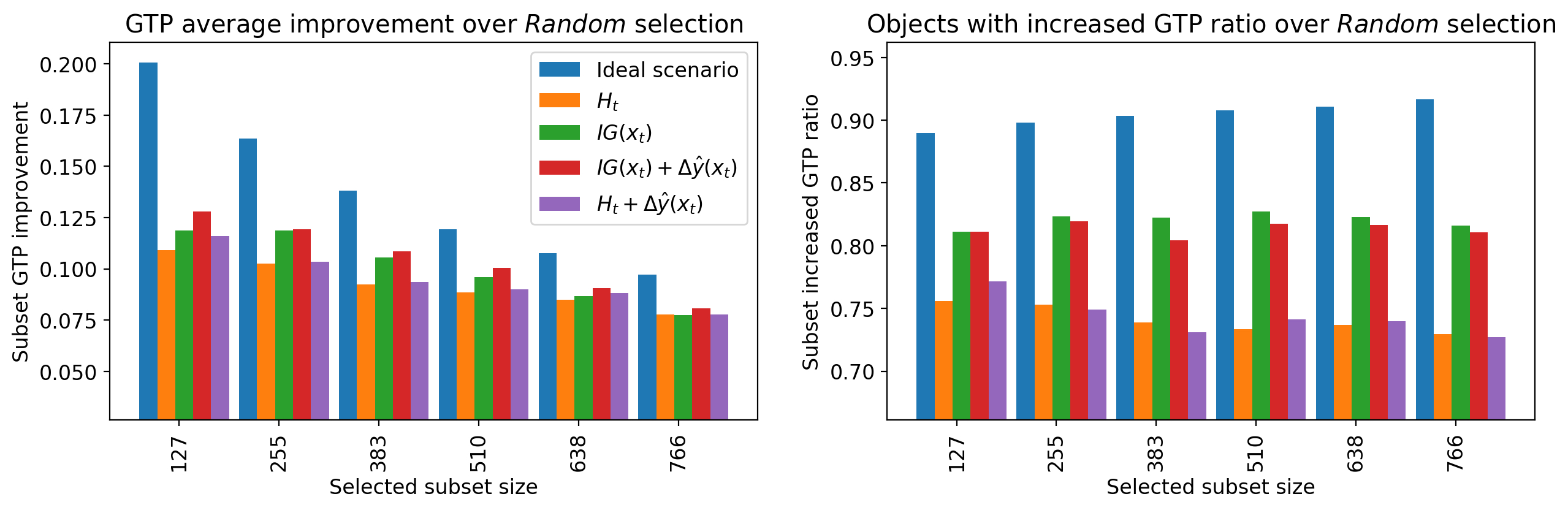}
\caption{\textbf{(Left)} Selected subset Ground Truth Probability (GTP) average improvement and \textbf{(Right)} ratio of objects in selected subset with increased GTP for different strategies and subset sizes, over $Random$ selection strategy.}
\label{figure:gtp}
\end{figure}

We note that it is more important to have a good performance at low $s$ values rather than at high $s$ values since we aim to have the largest improvement in classification prediction with the least querying for spectra. A close-up of the left plot in Figure \ref{figure:comparison} for the lower values of $s$ is shown in the left plot of Figure \ref{figure:gtp}. Here it is shown that $IG(x_t) + \Delta \hat{y}(x_t)$ has higher values for GTP mean improvement and also a higher gap with the base-line strategies for small $s$ values compared to high $s$ values. Take for example $s=127$ which represents 5\% of the dataset. If the objects of this subset are selected with $IG(x_t) + \Delta \hat{y}(x_t)$, $H_t$ and \textit{Random} strategies it reaches a GTP mean improvement of $0.13$, $0.11$ and $0.03$, respectively. This means that $IG(x_t) + \Delta \hat{y}(x_t)$ improves the GTP $1.18$ times as much as $H_t$ strategy and $4.33$ times as much as \textit{Random} strategy. The same example but with $s=510$ (20\% of the dataset) gives that $IG(x_t) + \Delta \hat{y}(x_t)$ improves the GTP of each selected object in average $1.11$ times as much as $H_t$ strategy and $3.33$ times as much as \textit{Random} strategy.

\begin{figure}[h]
\centering
\includegraphics[width=0.5\textwidth]{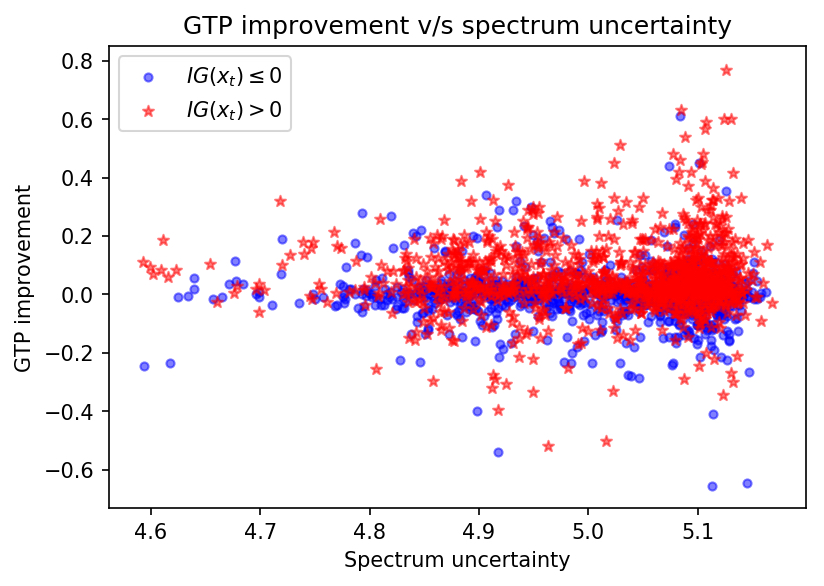}
\caption{Ground Truth Probability (GTP) improvement v/s spectrum uncertainty. Uncertainty is assessed as the entropy of the sampled spectra for each object in the dataset. Objects with positive approximate \textit{information gain} (eq. \ref{eqn:info_diff_update}) are plotted in red stars while non-positive objects are plotted in blue dots.}
\label{figure:uncertainty}
\end{figure}

These results show that using estimated spectra for object selection leads to better results compared to not using them. Strategies that use $IG(x_t)$ could be further improved to be as good as \textit{Ideal scenario} if we improve $Q(x_s|x_t)$ so that the estimation of the spectra is closer to the real ones. Objects for which we can achieve good estimations of their spectra may not be worth for spectra querying and hence we can save observational resources by not observing them. This is depicted in Figure \ref{figure:uncertainty} which shows the GTP improvement vs spectrum uncertainty for each object. To calculate spectrum uncertainty for a given object with $x_t$ features we first sample the most likely spectra from the conditional distribution $Q(x_s|x_t)$ which is a GMM as described in subsection \ref{subsection:joint_distribution}. Each sampled spectrum $\hat{x}_s$ has an assigned probability proportional to $Q(x_s=\hat{x}_s|x_t)$. Uncertainty then is assessed as the entropy of the sampled spectra $\hat{x}_{s,i} \sim Q(x_s|x_t)$, $i \in [1...N]$ for each object in the dataset. We focus on objects with positive approximate \textit{information gain} (eq. \ref{eqn:info_diff_update}), \text{i.e.} objects that have a higher priority assigned with the $IG(x_t)$ strategy. If the uncertainty is high the GTP improvement has a wide range of positive and negative values. While if the uncertainty is low the most likely the GTP improvement to be near 0 (no gain). Whenever we are fairly certain about an object's spectrum the less likely to gain improvement in classification results and thus not worth querying for it. On the other hand, if we are widely uncertain of an object's spectrum, we may gain a significant improvement in classification results if we query for its spectrum. This way it is worth improving our estimation of the spectra ($Q(x_s|x_t)$) so we can further save observational resources by not choosing objects for which we are already certain of their spectra outcome and instead choose the ones for which we are uncertain of their spectra outcome.

Complementary to the GTP average improvement over the selected subset, the right plot of Figure \ref{figure:gtp} shows the proportion of objects in the selected subset which improve their GTP in any positive amount, disregarding the average improvement. The best results are reached with $IG(x_t)$ and $IG(x_t) + \Delta \hat{y}(x_t)$. After those, $H_t + \Delta \hat{y}(x_t)$ and $H_t$ stay competitive and finally \textit{Random} presents the worst performance.

The middle plot of Figure \ref{figure:comparison} shows the accuracy (\%) improvement of the classification of the selected subset (y-axis). If an arbitrary $s$ is taken and it has an accuracy improvement of $0.3$ with some strategy, it means that it has $30\%$ more objects in the selected subset which are correctly classified using spectrum features ($l_{ts}(x_t, x_s)=y$) compared to not using them ($l_{t}(x_t)=y$), for that strategy selection method. As mentioned earlier in this subsection, $\Delta \hat{y}(x_t)$ signals 113 samples to prioritize first in the selection and thus the performance of strategies which use this \textit{indicator} are affected by this only up to $s=113$. $H_t + \Delta \hat{y}(x_t)$ and $IG(x_t) + \Delta \hat{y}(x_t)$ are the best strategies for improving accuracy of the selected subset on low $s$ values ($s<510$). This is to be expected since $\Delta\hat{y}(x_t)$ causes to select first the objects which are most likely to change their classification and hence improve accuracy. Next follows $H_t$. For higher values of $s$, $\Delta \hat{y}(x_t)$ no longer affects the selection and any strategy which uses $H_t$ is the best strategy. $IG(x_t)$ performs worse than the previous strategies for all $s$ but is still better than the $Random$ strategy. The latter is to be expected since $IG(x_t)$ selects objects which will improve their classification confidence regardless of the accuracy improvement.

We note that \textit{Ideal scenario} is worse than most strategies regarding accuracy improvement even though it uses the real spectra instead of an estimation. To improve accuracy we must select objects that are falsely predicted - False Positives (FP) - with time series features but are correctly predicted if spectrum features added, regardless of the amount of improvement of their GTP. \textit{Ideal scenario} uses only the current value of \textit{information gain} $IG(x_t)$ which selects objects which most likely increase their GTP but not necessarily will change their classification when spectrum information added. This is the reason why \textit{Ideal scenario} is worse at selecting objects which will improve accuracy compared to strategies that use $\Delta \hat{y}(x_t)$.

\begin{figure}[h]
\centering
\includegraphics[width=0.5\textwidth]{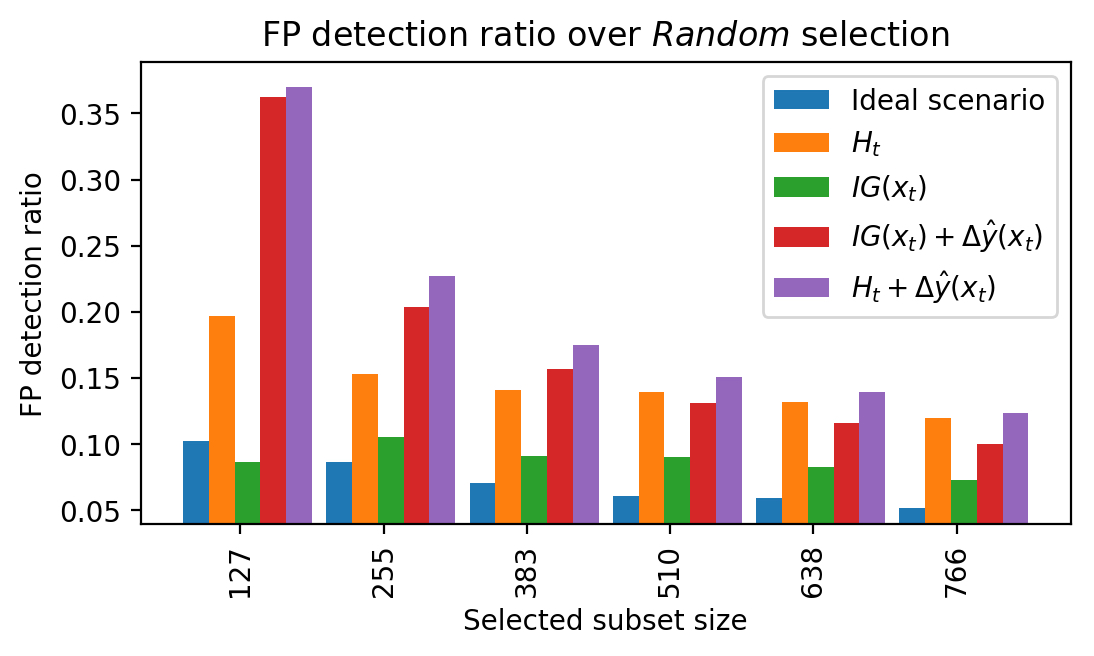}
\caption{Comparison of the selected subset ratio that change from a former falsely predicted - False Positives (FP) - class ($l_t(x_t) \neq y$) to their label $y$ ($l_{ts}(x_t, x_s) = y$), for different strategies and subset sizes, over \textit{Random} selection.}
\label{figure:fp}
\end{figure}
As mentioned before, $\Delta \hat{y}(x_t)$ returns the 113 ($4\%$ of the dataset) objects which are most probably to have a change of classification prediction if $x_s$ were queried. Within those objects, 46 change to their true class from a former falsely predicted class. The number of available falsely predicted - False Positives (FP) - objects in the dataset is equal to 390, from which 101 can correct their prediction if spectrum features are added. $\Delta \hat{y}(x_t)$ detects most of the FP which are correctable ($46\%$) with less than $~4\%$ of the dataset. This means that $\Delta \hat{y}(x_t)$ is a good detector of wrongly classified objects so that strategies which use it can quickly detect and correct FP objects (at least up until the number of objects with positive $\Delta \hat{y}(x_t)$).

The described FP object detection is depicted in Figure \ref{figure:fp} which shows the proportion of objects from the selected subset which change from a former falsely predicted class ($l_t(x_t) \neq y$) to their label $y$ ($l_{ts}(x_t, x_s) = y$). All strategies shows their results over \textit{Random} selection which has a constant FP detection ratio of 4$\%$. Our method $H_t + \Delta \hat{y}(x_t)$ has higher values for FP detection ratio and also a higher gap with the other strategies for small $s$ values compared to high $s$ values. For $s=127$ (5\% of the dataset), if the objects of this subset are selected with $H_t + \Delta \hat{y}(x_t)$, $H_t$ and \textit{Random} strategies, it corrects the classification of $37\%$, $20\%$ and $4\%$ of the selected objects, respectively. This means that $H_t + \Delta \hat{y}(x_t)$ corrects the classification of $17\%$ and $33\%$ more objects compared to $H_t$ and \textit{Random} strategies correspondingly. For $s=510$ (20\% of the dataset) we see that $H_t + \Delta \hat{y}(x_t)$ corrects the classification of $1\%$ (20) and $11\%$ more objects compared to $H_t$ and \textit{Random} strategies, respectively. 

From Figure \ref{figure:comparison} and \ref{figure:fp} we notice that for high $s$ values strategies which use $H_t$ are better detecting FP objects than strategies that do not. This means that after $\Delta \hat{y}(x_t)$ indicator, which covers low $s$ values, the most important selection criteria is the entropy of the time series classification outcome $H_{t}(\hat{y}| x_t)$. This is more clearly depicted in the right plot of Figure \ref{figure:comparison} which shows the overall accuracy (\%) of the classification of the dataset for different subset sizes and strategies. For any arbitrary strategy and subset size, the objects which are selected are classified using time series and spectrum features ($l_{ts}(x_t, x_s)$) while the non-selected objects are classified just with time series features ($l_{t}(x_t)$). Our $IG(x_t) + \Delta \hat{y}(x_t)$ and $H_t + \Delta \hat{y}(x_t)$ strategies reach same accuracies as $H_t$ and \textit{Random} strategies with smaller sized selected subsets, for all accuracies below $0.862$. To reach an accuracy of $0.86$, $H_t + \Delta \hat{y}(x_t)$, $H_t$ and \textit{Random} strategies need to query for the spectra of 101, 374 and 1315 objects correspondingly. $H_t + \Delta \hat{y}(x_t)$ needs 27\% and 8\% the amount of objects the base-lines strategies need to reach the same accuracy. 

The presented results suggest that the best strategy to use depends on the task it is meant for. If the aim is to improve the prediction probability (\textit{confidence}) then $IG(x_t) + \Delta \hat{y}(x_t)$ is the best choice, while if it is to detect false positives then any method which uses $\Delta \hat{y}(x_t)$ is the best choice. In both cases our methodologies surpasses base-line strategies ($H_t$ and \textit{Random}), specially within low subset sizes.

\section{CONCLUSIONS}
\label{section:conclusions}

In this paper, we develop a general method to select astronomical objects for which taking their spectrum would improve our knowledge regarding their classification. Adding spectra information provides further insights to time series information but requires more observational resources. Given current and future development of wide-field surveys such as LSST \citep{Ivezic2008}, it is valuable to know which objects should we prioritize to have spectrum in addition to time series given only a few spectroscopic facilities at hand. Differently from other works such as \cite{2019Ishida}, we make use of full spectral information through automatic spectrum feature extraction instead of querying for labels. Additionally, we do multiclass classification of objects as opposed to related works such as \cite{peters2015} and \cite{2019Ishida} which do binary classification. As a by-product of our work, we develop a model for the estimation of the spectrum of an object from its time series which may be used in other applications. To validate our method we perform extensive tests using a cross-match between spectra from SDSS DR14 \citep{Blanton2017} and CSS DR1 \citep{Drake2014}. The cross-matched catalog is provided  \href{https://drive.google.com/drive/folders/1AVentdOhgknlfCAz8aWOOm_fLU3BuosS}{here}. 

We propose multiple selection strategies based on two metrics. The first metric is $IG(x_t)$ (\textit{information gain}) which gives higher selection priority to the objects that are likely to improve their classification confidence if their spectra are queried. The second metric is $\Delta \hat{y}(x_t)$ which prioritizes the objects that are likely to change their classification if their spectra are queried. This metric ($\Delta \hat{y}(x_t)$) uses spectrum estimations to indicate the objects which will most likely change its classification. If the estimation is close enough to the real spectrum then it is reasonable that if the object changes its classification with the spectrum estimations then it will most probably change with the real spectrum too. We compare our strategies mainly with $H_t$ strategy which gives high priority to most uncertain objects on the time series classification outcome, similar to common active learning strategies such as \cite{2019Ishida}. Lastly, we also build strategies which mix $H_t$ and our metrics for object selection.

From the results, $\Delta \hat{y}(x_t)$ detects most of the falsely predicted - False Positives (FP) - which are correctable ($46\%$) with less than $~4\%$ of the dataset. This means that $\Delta \hat{y}(x_t)$ is a good detector of wrongly classified objects so that strategies which use it can quickly detect and correct FP objects. Subsets of candidates selected using $\Delta \hat{y}(x_t)$ have a higher improvement on classification accuracy compared to all other strategies, especially when a small number of objects are selected for spectrum follow-up. If a higher number of objects are selected then $\Delta \hat{y}(x_t)$ no longer affects the selection and any strategy that uses $H_t$ will be the best selection strategy. Subsets of candidates selected using $IG(x_t)$ have a higher improvement on the ground truth probability (GTP) (probability assigned to the \textit{real} class) compared to baseline strategies. This suggests that spectra querying may be used further from labelling to improve classification confidence of selected objects and more broadly the knowledge of those objects. Our developed \textit{information gain} $IG(x_t)$ metric leads us to select objects which are prone to improve their GTP and avoids objects for which we are fairly certain of their spectrum outcome and are not likely to gain improvement in classification results. This metric can be further enhanced if the estimation of the spectrum from the time series gets closer to the real spectrum. This way we could further save observational resources by not choosing objects for which we are already certain of their spectra outcome.

As future work, improvement of $x_t$ features with unsupervised time series feature learning could be included. Our method could be adapted for online selection so that the $Q(x_s|x_t)$ and $P_{ts}$ may be trained alongside with the choice of newly selected objects. The joint space of time series features and spectra features can be modeled otherwise so that the estimation of the spectrum gets closer to the real spectrum.

Additionally, automatic determination of the optimum number of objects to be queried could be developed so that when no significant improvement over any metric (accuracy and GTP for this case) is achieved, then no more spectra queries are done. Further metrics could be included to evaluate the improvement of our knowledge of the objects. 

Different alternatives within Design of Experiment may be explored. For example, finding the minimum spectrum resolution needed that still adds information to the time series. Alternatively, multiple options for enhancing the available information of an object could be included (simultaneously). A couple of examples of such are adding more points to the time series, changing the exposure time of observations or adding more color bands.

Finally, our methodology may be tested with other cross-matched catalogs such as \textit{Global Astrometric Interferometer for Astrophysics} \citep[GAIA,][]{GAIA2018} and SDSS \citep{york2000}. Additionally, we could apply our work to the recent Photometric LSST Astronomical Time Series Classification Challenge \citep[PLAsTiCC,][]{Hlozek2019} to rank objects for spectrum follow-up and compare with current works related to it.

\acknowledgments
We acknowledge the support from CONICYT-Chile, through the FONDECYT Regular projects 1180054 and 1170305 and from the Chilean Ministry of Economy, Development, and Tourism's Millennium Science Initiative through grant IC12009, awarded to The Millennium Institute of Astrophysics. Also, this research is supported by the Computer Science Department at PUC Chile, through the Fond-DCC project.

\appendix

\section{Appendix material}

\subsection{Variational Autoencoder Network}
\label{appendix: VAE}

Here we depict the selected Variational Autoencoder (VAE) for spectrum feature extraction, implemented with a neural network architecture. The preprocessing includes the normalization of the flux of each spectrum independently to $[0,1]$ range and the pruning of wavelengths in the tails so as to set a common wavelength grid of size equal to 3,794. The selected model consists of four encoding fully-connected layers with ReLU \citep{Glorot2010} activations with 2,847, 1,900, 953 and seven units respectively. Symmetrically, the decoder has four layers with an output layer size of 3,794 units and a sigmoid activation at the output \footnote{It is suitable to use sigmoid for this case since inputs are normalized to $[0,1]$ range.}. 

\begin{figure}
\centering
\includegraphics[width=0.4\textwidth]{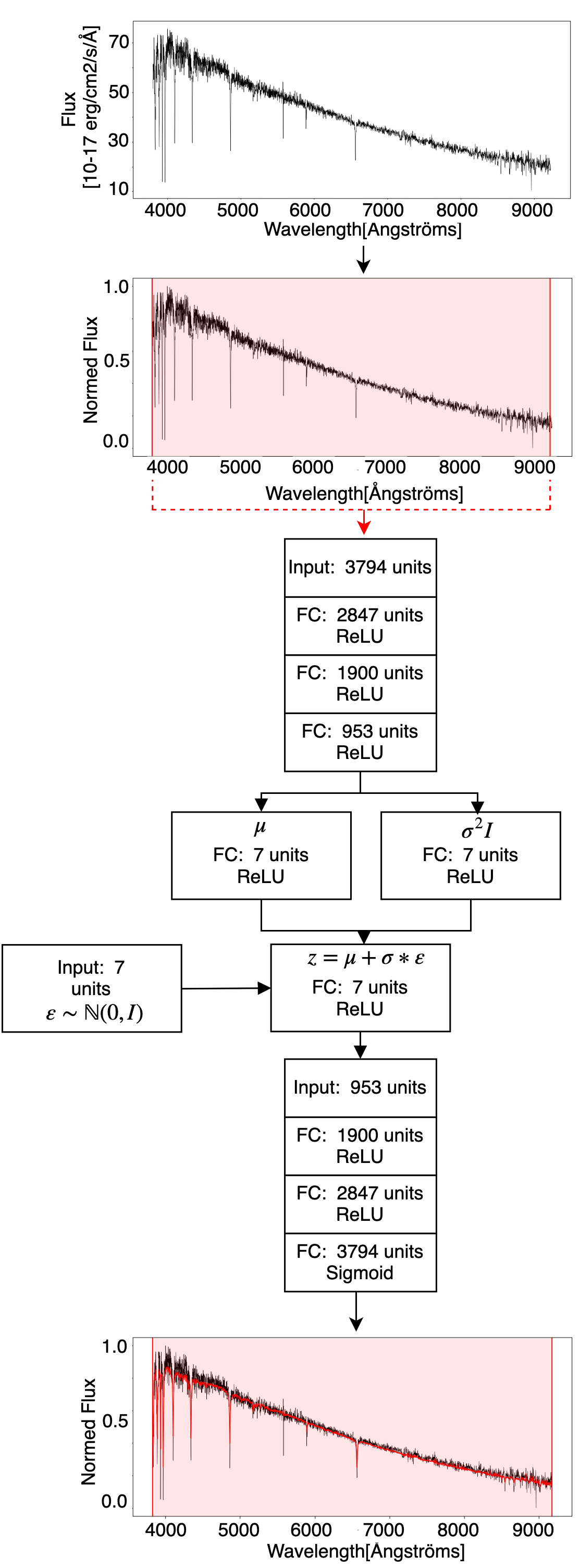}
\caption{Selected Variational Autoencoder network diagram ($z \in \mathbb{R}^{7}$). At the output, the red curve corresponds to the output of the VAE (decoded spectrum) while the black one corresponds to the original spectrum.}
\label{figure:vae_network}
\end{figure}

\newpage
\bibliography{references}


\end{document}